\documentclass[aps,prc,reprint,groupedaddress,amsmath,amssymb,showpacs,
showkeys,superscriptaddress,floatfix,nofootinbib]{revtex4-1}
\usepackage{graphicx}
\usepackage{dcolumn}
\usepackage{bm}
\begin{document}
\title{Probing of compact baryonic configurations in nuclei in $A(p,{\bar p})X$ reactions and antiproton
 formation length in nuclear matter}
\author{Yu. T. Kiselev}
\altaffiliation{Corresponding author. email: \it{yurikis@itep.ru}}
\author{V. A. Sheinkman}
\author{A. V. Akindinov}
\author{M. M. Chumakov}
\author{A. N. Martemyanov}
\author{V. A. Smirnitsky}
\author{Yu. V. Terekhov}
\affiliation {Institute for Theoretical and Experimental Physics,\\
Moscow 117218, Russia}
\author{E. Ya. Paryev}
\affiliation{Institute for Nuclear Research, Russian Academy of
Sciences,
\\ Moscow 117312, Russia}


\date{\today}

\begin{abstract}
Inclusive cross sections $\sigma^A=Ed^3{\sigma(X,P_t^2)/d^3p}$ of
antiproton and negative pion production on Be, Al, Cu and Ta targets hit
by 10~GeV protons were measured at the laboratory angles of
10.5$^{\circ}$ and 59$^{\circ}$. Antiproton cross sections were obtained
in both kinematically allowed and kinematically forbidden regions for
antiproton production on a free nucleon. The antiproton cross section
ratio as a function of the longitudinal variable $X$ exhibits three
separate plateaus which gives evidence for the existence of compact
baryon configurations in nuclei---small-distance scaled objects of
nuclear structure. Comparability of the measured cross section ratios
with those obtained in the inclusive electron scattering off nuclei
suggests a weak antiproton absorption in nuclei. Observed behavior of the
cross section ratios is interpreted in the framework of a model
considering the hadron production as a fragmentation of quarks
(antiquarks) into hadrons. It has been established that the antiproton
formation length in nuclear matter can reach the magnitude of 4.5~fm.
\end{abstract}

\maketitle

\section{\label{sec1}Introduction}

The normal nuclear density of 0.16--0.17~nucleon/fm$^3$ corresponds to
the average inter-nucleon (center-to-center) distance of 1.8--2~fm.
Comparing this value with the electromagnetic radius of a proton
$r_{em}=0.83$~fm shows that nucleons are tightly packed inside nuclei and
almost overlap. Due to quantum fluctuations of nuclear density, two or
more nucleons can get even closer, forming dense cold Compact Baryonic
Configurations (CBCs). Current estimates of the CBC size vary from 0.65
to 1~fm \cite{1}, \cite{2}, \cite{3} which corresponds to the density of
about four- to eightfold normal nuclear density. These values are
comparable with those expected in the cores of neutron stars. In the
conventional nucleon-meson nuclear physics, CBCs are described as
collections of closely packed nucleons usually referred to as Short Range
Correlations (SRCs) of nucleons \cite{4}. This description results in a
universal shape of the nuclear wave function for all nuclei at $k>k_F$,
where $k_F\sim$ 250~MeV/$c$ is the Fermi momentum of nucleon \cite{5}.
However, one can expect that the densities of CBCs are high enough to
modify the structure of underlying nucleon constituents. At short inter-
nucleon distances, nucleons can lose their identities and form multiquark
configurations \cite{3}, \cite{6}, \cite{7}, or multinucleon quark
clusters \cite{8}. Studying the CBCs provides information on the features
of the nuclear structure at small-distance scales as well as on the
equation of state of a cold and dense matter.

High energy lepton scattering on nuclei is an effective method of probing
CBCs. In deep inelastic physics, the cross sections are typically plotted
as functions of the Bjorken scaling variable $x_B = Q^2/2m\omega$, where
$Q$ is the four-momentum transferred to the system, $\omega$ is the
energy transfer, and $m$ is the mass of a proton. The value of $x_B$
determines the fraction of the struck quark's momentum relative to the
nucleon momentum in the infinite momentum frame. Obviously, a constituent
of an isolated nucleon carries $x_B \le 1$. Both approaches, SRC \cite{4}
and quark clustering in nuclei \cite{8}, lead to a simple prediction for
the region $x_B > 1$ where cooperative effects are required to provide
scattering process. Regardless of the nucleon or quark content of the
CBCs, the ratio of inclusive electron scattering cross sections for two
different nuclei has a scaling behavior at high-momentum transfer and
large $x_B$ values, which depends only on the ratio of probabilities to
find CBC in those two nuclei. This scaling manifests itself as a plateau
in the ratio when it is plotted as a function of $x_B$. Such plateaus
were first observed at SLAC \cite{9} and subsequently at Jefferson
Laboratory \cite{10,11}. In 1983, the European Muon Collaboration
measured the deep-inelastic per nucleon cross section ratio of $^{56}$Fe
over deuterium in a broad kinematic range \cite{12}. The measured ratio
revealed an unexpected structure which was subsequently confirmed by SLAC
\cite{13} and became known as the ``EMC effect''. Plotted as functions of
the Bjorken scaling variable $x_B$, the ratios showed a clear decrease at
$0.3<x_B<0.7$ and sharp rise at $x_B>0.8$. These results generated
considerable experimental and theoretical interest, but none of the
existing models has been able to explain the effect over the whole range
of $x_B$ and $A$. New high-precision experimental data on the EMC effect
in light nuclei obtained in Jefferson Laboratory \cite{14} suggest that
the slope in the $0.3 < x_B < 0.7$ region is an effect of local density
and is not a bulk property of the nuclear medium. At similar four-
momentum transfers, data from Hall C \cite{11} indicated that the nuclear
scaling plateaus at $x_B>1$ also represent an effect of local density.
The authors of \cite{15} found a linear correlation between the strength
of the EMC effect, defined as the slope of the cross section ratios in
the range $0.3<x_B<0.7$, and the nuclear scaling plateaus at $x_B > 1$.
These results are consistent with the idea that both effects scale with
the local nuclear environment.

Studying the cumulative reactions provides a complementary information on
the localised dense objects inside nuclei. Cumulative hadron production
can be the key to understanding the process of hadron creation in the
fragmentation region of a nucleus beyond the kinematical limits of the
production of these hadrons in the collisions of elementary projectiles
with isolated nucleon. In 1980--1990s the cumulative hadron production
was intensively investigated in many experiments using a variety of beam
particles and energies. The proton induced production of high momentum
cumulative protons, pions, kaons, antikaons and antiprotons \cite{2},
\cite{16} has been studied up to the value of $X \sim 3.5$, where $X$ is
a variable similar to the Bjorken $x_B$ (see Section \ref{sec3}). The
uncertainty principle requires that such hadrons may arise from
fluctuations of dense CBCs consisting of either a few closely packed
nucleons with large relative momentum \cite{4}, or from quark
constituents of a multiquark configuration carrying a light cone momentum
fraction greater than that of a single nucleon \cite{3}. It is worth
noting that the study of superfast quarks in deep inelastic scattering at
$x_B>1$ is one of the three key experimental programs of the future
12~GeV JLab facility \cite{1}. In the cumulative reactions, a nucleus
acts both as a target and an analyzer of the interaction of the products
with the nuclear environment. Therefore, the study of cumulative
reactions provides information on the hadronization process in nuclear
matter, a subject which is also of great current interest. Investigation
of cumulative antiproton production in $pA$ reactions is of special
interest since antiproton does not contain nuclear valence quarks. In
this paper we report on the study of the cumulative antiproton production
off nuclei. Two experiments were performed using the internal 10~GeV beam
of the ITEP proton synchrotron and nuclear targets. At the first
experiment the invariant cross sections were measured at 59$^{\circ}$
(lab) in the antiproton momentum interval of 0.6--1.7 GeV/$c$ which
corresponds to the cumulative region. At the second experiment the data
on the cross sections were obtained at 10$^{\circ}$ (lab) in the momentum
range allowed for the antiproton production on free nucleon. Invariant
cross sections of $\pi^-$ meson production on the same set of nuclear
targets were measured simultaneously. The goals of the experiments were
to search for the presence of CBCs in nuclei as well as to study the
antiproton absorption inside the nuclear matter.

\section{\label{sec2}Experimental arrangement}

The experiments were carried out with the internal proton beam of
the 10~GeV ITEP synchrotron irradiating Be, Al, Cu and Ta strip
targets, 60-- 150 micron thick. Secondary particles produced at
10.5$^{\circ}$ and 59$^{\circ}$ (lab) in the momentum range from 0.6
to 2.5 GeV/$c$ were detected by two identical arms of the Focusing
Hadron Spectrometer (FHS). Each arm was composed of two bending
dipole magnets and two pairs of quadrupole magnets. The momentum and
angular acceptance of the magnetic channels were ${\Delta}p/p = 1$\%
and 0.8 msr, correspondingly. Two multiwire proportional chambers
located at the second focus of each magnetic system controlled the
beam position during the data taking. The particle identification
system included a two-stage time-of-flight (TOF) system based on
scintillation counters and two Cherenkov counters. TOF measurements
were performed over two bases of 11 and 17 meters. Two
photomultipliers XP2020 mounted on both sides of each scintillator
provided the mean-time signal for TOF measurement and two energy
loss (d$E/$d$x$) signals. The time resolution of the TOF system was
less than 300~ps (FWHM). In the momentum range 0.6--1.3~GeV/$c$, the
threshold Cherenkov counter with water radiator was used for the
suppression of pion background. Analysis of the information on
d$E/$d$x$ and TOF allowed selection of antiprotons in the above
mentioned momentum range. At higher ${\bar p}$ momenta, the trigger
also included a signal from the differential Cherenkov counter. The
counter optics was adjusted for detection of antiprotons with given
momentum selected by the magnetic system. Photons emitted by
antiprotons in the radiator of the differential Cherenkov counter
(glycerin-water mixture) were detected by the ring consisting of 12
photomultipliers (PMs). The velocity resolution of Cherenkov counter
was ${\Delta}\beta/\beta = 2$\%. The antiproton selection criterion
handled by hardware included a requirement that 8 PMs of the ring
were fired in a single event. For each event, the information on
larger number of fired photomultipliers (9 and 10) was also
recorded. The identification reliability increased with the
increasing number of operated PMs, however leading to a decrease of
the detection efficiency. This efficiency varied from 35\% to 75\%
depending on the number of operated PMs and was periodically tested
in special measurements using protons with the same velocity.
Depending on the effect-to-background ratio, the information on
different number of fired photomultipliers was used during the
off-line analysis. Selection of antiprotons was quite reliable down
to the ratio of ${\bar p}/\pi^-=10^{-6}$. The misidentification
probability was less than 4\% for almost all data and did not exceed
7\% at the lowest cross section value. Secondary antiprotons and
pions were detected simultaneously. The measured antiproton and
negative pion yields were corrected for the loss due to the nuclear
interactions and multiple scattering in the material of
spectrometer, detector efficiencies and pion in-flight decay at the
distance of 32 meters from the production target towards the second
focus of the magnetic channels. In order to determine the absolute
value of the pion differential cross sections on different nuclear
targets, the $\pi^- $ yields at 10$^{\circ}$ and 59$^{\circ}$ were
measured along with the flux of protons traversing the sandwich type
targets made of thin (12 micron) Al foil and Be, Al, Cu or Ta strips
used as the production targets. Determination of the incident proton
flux was performed by measuring the induced $\gamma$-activity in the
reaction $p+^{27}{\rm Al}\rightarrow^{24}{\rm Na^{*}} + X$. Knowing
the cross section of this reaction and the acceptance of the
magnetic channels allowed calculation of the differential cross
sections for $\pi^-$ production at 10$^{\circ}$ and 59$^{\circ}$.
Pion cross sections were then used for the absolute normalization of
the differential cross sections for antiproton production since the
ratio of the antiproton yield to that of the pion was measured at
each secondary momentum. The resulting uncertainty of the absolute
normalization of the cross sections was estimated as 17\% and 20\%
for the data obtained at 10$^{\circ}$ and 59$^{\circ}$,
correspondingly. The measured invariant cross sections are presented
in Tables~\ref{tab:table1}--\ref{tab:table4}. The errors quoted in
these Tables do not include uncertainties in the absolute
normalization.

\begin{table*}
\begin{ruledtabular}
\caption{\label{tab:table1}Invariant cross sections
$\sigma^A=Ed^3{\sigma/d^3p}$ [GeV\mbox{{$\mu$}b}/(GeV/c)$^3$] of ${\bar
p}$ production at the total energy of the incident proton of 10~GeV and
at the lab angle of 10.5$^{\circ}$. Statistical (first) and systematical
(second) errors are presented. The systematic errors include the
correction uncertainties and reproducibility of the measurements.}
\begin{tabular}{|c|c|c|c|c|}\hline
 $p$ (GeV/$c$)& Be & Al & Cu & Ta \\
\hline
0.73 & 6.48$\pm$0.32$\pm$0.8 & 18.1$\pm$0.9$\pm$2.2 &
28.3$\pm$1.4$\pm$3.4 & 42.6$\pm$2.1$\pm$5.1 \\
\hline
0.93 & 11.6$\pm$0.46$\pm$1.2 & 31.1$\pm$1.2$\pm$3.1 &
45.0$\pm$1.8$\pm$4.5 & 58.8$\pm$2.4$\pm$5.9 \\
\hline
1.31 & 23.0$\pm$0.98$\pm$2.5 & 61$\pm$2.6$\pm$6.5 & 91.1$\pm$3.9$\pm$9.7
& 141$\pm$6$\pm$15 \\
\hline
1.76 & 19.3$\pm$0.78$\pm$1.5 & 46.7$\pm$1.9$\pm$3.7 &
66.6$\pm$2.7$\pm$5.3 & 111$\pm$4.4$\pm$8.9 \\
\hline
2.47 & 8.51$\pm$0.34$\pm$0.7 & 20.4$\pm$0.82$\pm$1.6 &
27.7$\pm$1.1$\pm$2.2 & 47.3$\pm$1.9$\pm$3.8 \\
\hline
\end{tabular}
\end{ruledtabular}
\end{table*}

\begin{table}
\caption{\label{tab:table2}Invariant cross sections
$\sigma^A=Ed^3{\sigma/d^3p}$ [GeV\mbox{mb}/(GeV/c)$^3$] of $\pi^-$
production at the total energy of the incident proton of 10~GeV and at
the lab angle of 10.5$^{\circ}$. Statistical errors are negligible. The
systematic error of each cross section equals to 8\% and includes the
uncertainties of the corrections and reproducibility of the
measurements.}
\begin{ruledtabular}
\begin{tabular} {|c|c|c|c|c|}\hline
 $p$ (GeV/$c$)& Be & Al & Cu & Ta \\
\hline
0.73 & 300 & 762 & 1214 & 2198 \\
\hline
0.93 & 194 & 483 & 736 & 1286 \\
\hline
1.31 & 96.8 & 233 & 351 & 603 \\
\hline
1.76 & 45.6 & 108 & 160 & 268 \\
\hline
2.47 & 15.8 & 36.7 & 53.1 & 88.7 \\
\hline
\end{tabular}
\end{ruledtabular}
\end{table}

\begin{table*}
\caption{\label{tab:table3}Invariant cross sections
$\sigma^A=Ed^3{\sigma/d^3p}$ [GeV\mbox{nb}/(GeV/c)$^3$] of ${\bar p}$
production at the total energy of the incident proton of 1~ GeV and at
the lab angle of 59$^{\circ}$. Statistical (first) and systematical
(second) errors are presented. Systematic errors include the correction
uncertainties and reproducibility of the measurements.}
\begin{ruledtabular}
\begin{tabular} {|c|c|c|c|c|}\hline
 $p$ (GeV/$c$)& Be & Al & Cu & Ta \\
\hline
0.58 & 61$\pm$4$\pm$7 & 216$\pm$15$\pm$26 & 472$\pm$33$\pm$57 &
634$\pm$44$\pm$76 \\
\hline
0.69 & 82$\pm$5$\pm$9 & 367$\pm$22$\pm$40 & 718$\pm$43$\pm$79 &
1129$\pm$68$\pm$124 \\
\hline
0.78 & 78$\pm$5$\pm$8 & 329$\pm$20$\pm$33 & 614$\pm$37$\pm$61 &
967$\pm$58$\pm$97 \\
\hline
0.88 & 79$\pm$5$\pm$8 & 331$\pm$20$\pm$33 & 681$\pm$41$\pm$68 &
938$\pm$56$\pm$94 \\
\hline
0.98 & 51$\pm$3$\pm$5 & 260$\pm$13$\pm$26 & 408$\pm$20$\pm$41 &
744$\pm$37$\pm$74 \\
\hline
1.07 & 38.1$\pm$1.5$\pm$3.8 & 178$\pm$7$\pm$18 & 234$\pm$9$\pm$23 &
456$\pm$22$\pm$46 \\
\hline
1.19 & 14.0$\pm$0.6$\pm$1.3 & 85$\pm$3$\pm$8 & 172$\pm$7$\pm$15 &
299$\pm$14$\pm$27 \\
\hline
1.35 & 5.7$\pm$0.3$\pm$0.5 & 32$\pm$2$\pm$3 & 79$\pm$5$\pm$6 &
140$\pm$10$\pm$11 \\
\hline
1.53 & 1.4$\pm$0.14$\pm$0.11 & 11.6$\pm$1.2$\pm$0.9 & 28$\pm$2.8$\pm$2.2
& 40$\pm$4$\pm$4 \\
\hline
1.72 & 0.3$\pm$0.04$\pm$0.02 & 2.5$\pm$0.35$\pm$0.2 & 5.1$\pm$0.7$\pm$0.4
& \\
\hline
\end{tabular}
\end{ruledtabular}
\end{table*}

\begin{table}
\caption{\label{tab:table4}Invariant cross sections
$\sigma^A=Ed^3{\sigma/d^3p}$ [GeV\mbox{mb}/(GeV/c)$^3$] of $\pi^-$
production at the total energy of the incident proton of 10~GeV and at
the lab angle of 59$^{\circ}$. Statistical errors are negligible.
Systematic error of each cross section is 8\% and includes the
uncertainties of
the corrections and reproducibility of the measurements.}
\begin{ruledtabular}
\begin{tabular} {|c|c|c|c|c|}\hline
 $p$ (GeV/$c$)& Be & Al & Cu & Ta \\
\hline
0.58 & 20.3 & 59.3 & 121 & 226 \\
\hline
0.69 & 12.6 & 36.7 & 70 & 134 \\
\hline
0.78 & 7.03 & 21.6 & 43.8 & 79.6 \\
\hline
0.88 & 4.02 & 11.6 & 24.1 & 45.3 \\
\hline
0.98 & 1.81 & 5.92 & 12.1 & 23.5 \\
\hline
1.07 & 0.87 & 3.03 & 6.33 & 12.2 \\
\hline
1.19 & 0.332 & 1.25 & 2.85 & 5.43 \\
\hline
1.35 & 8.21*$10^{-2}$ & 0.43 & 0.922 & 1.81 \\
\hline
1.53 & 2.04*$10^{-2}$ & 0.114 & 0.271 & 0.542 \\
\hline
1.72 & 3.02*$10^{-3}$ & 1.96*$10^{-2}$ & 4.97*$10^{-2}$ & 1.13*$10^{-1}$ \\
\hline
\end{tabular}
\end{ruledtabular}
\end{table}

\section{\label{sec3}Results and discussion}

Measured at the experiment were the invariant inclusive cross sections
$\sigma^A=Ed^3{\sigma(X,P_t^2)/d^3p}$ of antiproton production on target
nucleus with atomic number $A$. Transverse variable $P_t^2$ was derived
from the absolute value of antiproton momentum and from its production
angle. To determine longitudinal variable $X$ we used the minimal mass of
intranuclear target, expressed in the units of nucleon mass $m$, for
which the production of an antiproton with registered 3-momentum was
kinematically possible. The variable $X$ is called a cumulative number.
It can be obtained from Eq.~(\ref{eq:one}) expressing the conservation of
energy-momentum and baryonic number at the reaction $p+mX \to {\bar
p}+\ldots$:
\begin{equation}
\label{eq:one}
({\hat P}_0+m{\hat X}-{\hat P})^2 \ge [(2+X)m]^2,
\end{equation}
where ${\hat P}_0$, ${\hat P}$ and $m{\hat X}(mX,{\bf 0})$ are the 4-
momenta of the incident proton, detected antiproton and intranuclear
target, respectively. By equating the left- and right-hand sides of
Eq.~(\ref{eq:one}), one can obtain:
\begin{equation}
\label{eq:two}
X=\left(1-\frac{E}{E_0}-\frac{2m}{E_0}\right)^{-1}\left(\frac{E-
{\beta_0}P\cos{\theta}}{m}+\frac{m}{E_0}\right),
\end{equation}
where $E_0$ and $P_0$ are the total energy and 3-momentum of the
projectile proton, ${\beta}_0=P_0/E_0$; $E$, $P$ and $\theta$ are the
total energy, 3-momentum and antiproton production angle in the
laboratory system, respectively. The value of $X=1$ corresponds to the
kinematical limit of antiproton creation on nucleon at rest, while in the
cumulative region $X > 1$. The definition of $X$ (Eq.~(\ref{eq:two}))
takes into account the finite energy $E_0$ of the projectile proton. As
the energy $E_0$ increases, $X$ tends to the light cone variable
$\alpha=(E-P_{l})/m$. Note that the Bjorken variable $x_B$ can also be
interpreted as the minimum target mass in the nucleon mass units, i.e.
the variable expressing the number of nucleons involved in the deep
inelastic scattering process. The invariant inclusive cross section of
the antiproton production on nucleus $À$ can be represented in the form
similar to that used in \cite{10} for the description of inclusive
electron scattering on nuclei:
\begin{equation}
\label{eq:three}
\sigma^A(X,P_t^2)=A\sum_{j=1}^A\frac{W_j^A}{j}\sigma_j(X,P_t^2)f_{\rm
FSI}^A\theta(j-X),
\end{equation}
\begin{equation}
\label{eq:four}
\sum_{j=1}^A W_J^A=1.
\end{equation}
Here, $W_j^A=m_j/A$ is the per nucleon probability that $m_j$
nucleons of the target nucleus with the mass number $A$ belong to
CBCs with a given $j$, $f_{\rm FSI}^A$ stands for the factor
describing the absorption of the outgoing antiproton on its way out
of the nucleus, and $\theta(x)$ is the step function. Depending on
the model of cumulative hadron production, $\sigma_j(X,P_t^2)$ can
be considered either as a cross section of the proton–-$j$-nucleon
correlation \cite{4}, or as a cross section of the proton
interaction with the colorless configuration consisting of $j$ 3$q$
systems \cite{3}. Eq.~(\ref{eq:three}) does not take into account
the initial state interaction (ISI) of the projectile proton for the
reason discussed later in this Section. Since the probabilities
$W_j^A$ have to drop fast with $j$, one may expect that interaction
with $j$-CBC will dominate in the region $j-1 < X < j$. Therefore,
the ratio of the nuclear antiproton production cross sections per
nucleon of heavy $A_1$ and light $A_2$ nuclei in this region have to
be independent of the cross section $\sigma_j(X,P_t^2)$ and have
discrete values for different $j$:
\begin{equation}
\label{eq:five}
R_j=\frac{A_2}{A_1}\frac{\sigma^{A_1}}{\sigma^{A_2}}=\frac{W_j^{A_1}}{W_j
^{A_2}}\frac{f_{\rm FSI}^{A_1}} {f_{\rm FSI}^{A_2}}.
\end{equation}
Since the ratio $\frac{W_j^{A_1}}{W_j^{A_2}}$ has to grow with $A$ and
$j$, one can expect that discrete values of $R_j$ would also exhibit a
similar behavior provided that the ratios of the absorption factors
$\frac{f_{\rm FSI}^{A_1}}{f_{\rm FSI}^{A_2}}$ are almost constant at each
$j$.

The experimental data on inclusive electron scattering cross sections on
nuclei $^3$He, $^4$He, $^{12}$C, $^{56}$Fe obtained by the CLAS
Collaboration (JLab) \cite{10} at the four-momentum transfer to the
target $1.4 < Q^2 < 2.6$~(GeV/c)$^{2}$ and at the Bjorken variable $1 <
x_B < 2.8$ indicate the existence of such regions. Cross section ratios
$R=(3\sigma^A)/(A\sigma^{^3{\rm He}})$ exhibit two separate plateaus at
$1.5 < x_B < 2$ and at $x_B > 2.25$, interpreted by the authors of
\cite{10} as an evidence for the presence of compact 2-nucleon and 3-
nucleon SRCs in nuclei. Similar behavior of the cross section ratios was
observed later by the HMS Collaboration at JLAB \cite{11}.

\begin{figure*}
\includegraphics[width=19.0cm]{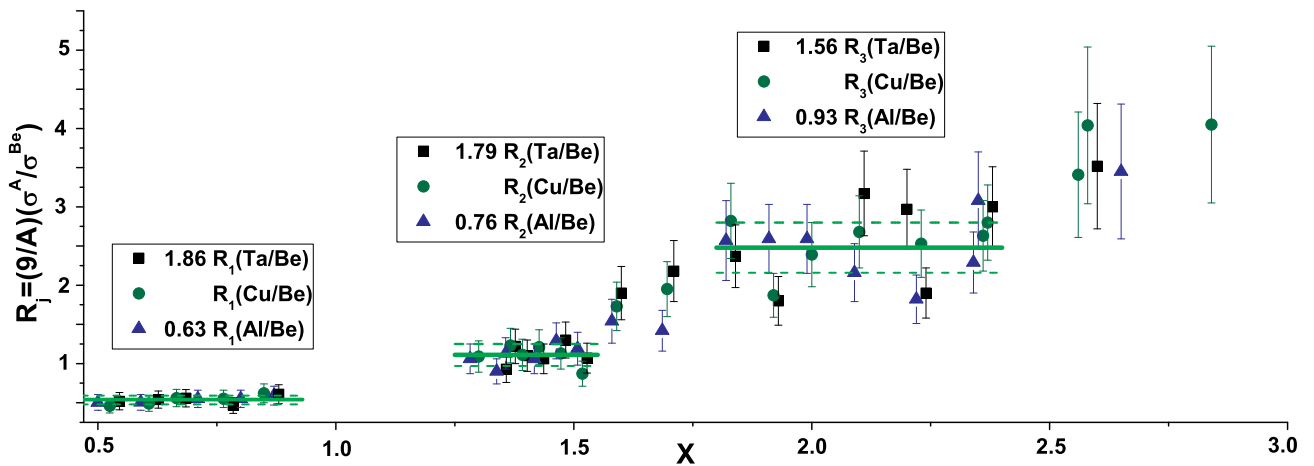}
\caption{\label{fig:fig1} (Color online) Cross section ratio
$R=(9\sigma_{pA \to {\bar p}})/(A\sigma_{p{\rm Be} \to {\bar p}})$
as a function of $X$ in the range $0.5 < X < 2.8$. Values of the
rescaling factors for $j=1, 2, 3$ are indicated in the legends from
left to right, respectively. Solid and dashed lines correspond to
the weighted average magnitudes of $R({\rm Cu}/{\rm Be})$ and their
errors shown in Table~\ref{tab:table5}. Symbols for Al/Be and Ta/Be
ratios are slightly displaced.}
\end{figure*}

In the subsequent analysis, along with the cross sections measured in the
present experiment at the lab angles of 10$^{\circ}$ and 59$^{\circ}$, we
also use data on cumulative antiproton production by 10~GeV protons
obtained on the same set of target nuclei at the lab angles of
97$^{\circ}$ and 119$^{\circ}$ which was presented in our previous work
\cite{16}. The $X$ dependence of the ratio $R_j(X)$ in the region $0.53 <
X < 2.83$ is shown in Fig.~\ref{fig:fig1}.

To analyse the behavior of $R_j(X)$ in more detail, the measured ratios
$R_j=(9\sigma^{\rm Al})/(27\sigma^{^{\rm Be}})=R_j({\rm Al}/{\rm Be})$
and $R_j=(9\sigma^{\rm Ta})/(181\sigma^{^{\rm Be}})=R_j({\rm Ta}/{\rm
Be})$ were multiplied by factors chosen in such a way that their average
values were equalized with the average magnitude of the ratio
$R_j=(9\sigma^{\rm Cu})/(64\sigma^{^{\rm Be}})=R_j({\rm Cu}/{\rm Be})$.
The first plateau is clearly seen at $0.5 < X < 0.85$, the second
plateau---at $1.3 < X < 1.6$. The ratio increases with $X$, which is
related to the contribution of CBCs with $j=3$. Fig.~\ref{fig:fig1} also
demonstrates the presence of a third plateau in the region $1.8 < X <
2.4$. The observed increase in the cross section ratio at $X > 2.5$ can
be explained as a transition from CBC with $j=3$ to CBC with $j=4$.

Note that the absolute values of antiproton production cross sections in
the plateau regions vary by several orders of magnitude. The plotted
cross section ratios correspond to different antiproton momenta and
production angles and, hence, to different values of the transverse
momenta $P_t^2$ varying from 0.3 to 2~(GeV/$c$)$^2$. Therefore, the
ratios $R_j({A}/{\rm Be})$ exhibit the scaling behavior on each plateau
($j=1,2,3$), i.e. they do not depend on $X$ and on $P_t^2$. Similar
behavior of the cross section ratios (independence from $x_B$ and $Q^2$)
was observed in JLab experiments \cite{10,11}. It should be noticed that
the plateaus, which correspond to $j=2,3$ in the JLab results
\cite{10,11} were observed at larger values of $x_B$ than those values of
$X$ obtained in our experiment. The main reason for this apparent
discrepancy is the use of different variables: $X$ is similar but not
equivalent to $x_B$. Besides, as discussed later in this Section, there
is a physical reason for the downward shift of the plateaus in the
reactions with detection of a hadron in the final state compared to
inclusive electron scattering reactions.

The cross section ratios can be determined more precisely than the
absolute values of cross sections since the detector-dependent errors,
many of the corrections factors and the most part of the absolute
normalization uncertainty cancel out in the ratios. As a result, the
error of each individual value of the cross section ratio $R_j({A}/{\rm
Be})$ shown in Fig.~\ref{fig:fig1} contains only a residual of the
systematic error. The error bars manifest statistical and remaining
systematic errors added in quadrature. The weighted average magnitudes of
$R_j({A}/{\rm Be})$ are presented in Table~\ref{tab:table5}.

\begin{table}
\caption{\label{tab:table5}Cross section ratio $R_{j}=(9\sigma_{pA \to
{\bar p}})/(A\sigma_{p{\rm Be} \to {\bar p}})$ for $j=1, 2, 3$.}
\begin{ruledtabular}
\begin{tabular}{|c|c|c|c|}\hline
 & & & \\
 $R_{j}$ & $j=1$ & $j=2$ & $j=3$ \\
 & ($0 \le X \le 1$) & ($1 \le X \le 2$) & ($2 \le X \le 3$)\\
\hline
Al/Be & 0.86$\pm$0.08 & 1.45$\pm$0.12 & 2.77$\pm$0.25 \\
\hline
Cu/Be & 0.54$\pm$0.05 & 1.11$\pm$0.09 & 2.48$\pm$0.22 \\
\hline
Ta/Be & 0.29$\pm$0.03 & 0.63$\pm$0.05 & 1.74$\pm$0.20 \\
\hline
\end{tabular}
\end{ruledtabular}
\end{table}

As seen from Eq.~(\ref{eq:five}), the values of $R_j(A/{\rm Be})$ depend
on the probabilities $W_j^A$. These probabilities were calculated within
a model identifying CBCs with SRCs \cite{4}, and within a model
considering CBCs as quark clusters \cite{8}. Calculations of the
probabilities $W_j^A$ for $j=2,3$ in \cite{4} are based on the analysis
of hadron-production on light nuclei and on the Fermi liquid theory of
$^{56}$Fe. According to the model \cite{8} two nucleons form a 6-quark
cluster when they are separated by less than a critical distance $d_c =
2R_c = 1$~fm, where $R_c$ is a nucleon critical ``color percolation''
radius. A third nucleon distanced less than $d_c$ from the first two may
join the cluster to form a 9-quark cluster, etc. In both approaches, the
calculation results are model-dependent. Since the values of $W_j^A$ for
$j=2,3$, obtained in the model \cite{4}, have essentially weaker $A$- and
$j$-dependencies compared to those calculated in \cite{17} within the
model \cite{8}, the absolute values of $W_2^A$ and $W_3^A$ in these
models differ significantly (see discussion in \cite{10}). However, the
probability ratios $W_j^A/W_j^{\rm Be}$, entering Eq.~(\ref{eq:five}),
almost do not depend on the chosen model. For example, the difference in
the magnitudes of ratios $W_2^{\rm Fe}/W_2^{\rm C}$ and $W_3^{\rm
Fe}/W_3^{\rm C}$ from \cite{4} and \cite{17} does not exceed 5\%. The
probabilities $W_j^A$ have been calculated in \cite{4} for light nuclei
and for the iron nucleus ($A=56$), whereas in \cite{17} they were
obtained at the same footing for a wide range of nuclei from helium to
uranium. Therefore, in the subsequent analysis we will use the
probability ratios presented in Table~\ref{tab:table6}, $W_j^A/W_j^{\rm
Be}$ ($j=1,2,3$) calculated in \cite{17} within the quark-cluster-model
\cite{8}.

\begin{table}
\caption{\label{tab:table6}Probability ratios $W_j^A/W_j^{\rm Be}$
calculated basing on the data from Table~\ref{tab:table4} \cite{17}.}
\begin{ruledtabular}
\begin{tabular} {|c|c|c|c|}\hline
 & & & \\
$W_j^A/W_j^{\rm Be}$ & $j=1$ & $j=2$ & $j=3$ \\
\hline
Al/Be & 0.92 & 1.65 & 2.8 \\
\hline
Cu/Be & 0.90 & 1.80 & 3.4 \\
\hline
Ta/Be & 0.88 & 1.86 & 3.6 \\
\hline
\end{tabular}
\end{ruledtabular}
\end{table}

Comparing the measured ratios $R_j({\rm Al}/{\rm Be})$ (the first line of
Table~\ref{tab:table5}) to the ratios $W_j^{\rm Al}/W_j^{\rm Be}$ (the
first line of Table~\ref{tab:table6}), one can see that the ratio of the
absorption factors $f_{\rm FSI}^{\rm Al}/f_{\rm FSI}^{\rm Be}$ in
Eq.~(\ref{eq:five}) is close to 1 for all $j$. This suggests a weak
absorption of antiprotons in light nuclei, up to Al. The lesser values of
the cross section ratios, given in the second and third lines of
Table~\ref{tab:table5}, when compared to the probability ratios presented
in the corresponding lines of Table~\ref{tab:table6}, indicate the
presence of an antiproton absorption in the intermediate and heavy
nuclei.

\begin{figure}
\includegraphics[width=8.0cm]{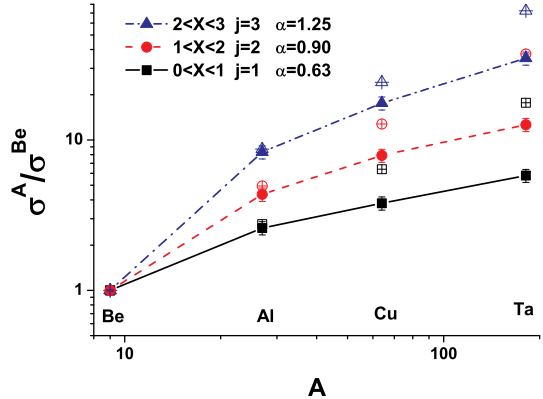}

\caption{\label{fig:fig2} (Color online) $A$-dependence of the cross
section ratio $\sigma^A/\sigma^{\rm Be}$ for $j=1, 2, 3$. Full
symbols are data points. The values of exponent $\alpha$ from the
power law approximation of the measured ratio $(A/9)^{\alpha}$ are
indicated in the legend. The magnitudes of the cross section ratio
calculated using the data of Table~\ref{tab:table6} are shown by
crossed empty symbols.}
\end{figure}

Fig.~\ref{fig:fig2} shows the $A$-dependence of the antiproton
production cross section ratios ${\tilde R}_j(A/{\rm Be})=R_j(A/{\rm
Be})(A/9)$ for various $j$. The points correspond to the ratios
$R_j(A/{\rm Be})$, presented in Table~\ref{tab:table5}, and they are
linked by lines to guide the eyes. One can see that this dependence
becomes steeper with the increase of $X$. The main reason for such
change of $A$-dependence is the fast growth of the ratios
$W_j^{A}/W_j^{\rm Be}$ with $j$. Despite the fact that the
experimental dependencies do not follow well the simple
one-parameter power law dependence $\sigma^A=\sigma_0A^{\alpha}$, we
calculate the exponent $\alpha$ to characterize its change with $j$.
The exponent $\alpha$, calculated with accounting for antiproton
creation cross sections on all nuclei in the region where CBCs with
$j=3$ dominate in the antiproton production, exceeds 1 and is
approximately equal to 1.25.\footnote{Similar values of $\alpha > 1$
were observed in the antiproton production on nuclei with large
transverse momentum (``Cronin effect'') \cite{18}. In both cases the
nucleons behave cooperatively.} Experimental value of the ratio
${\tilde R}_3({\rm Al}/{\rm Be}) =8.3$ corresponds to
$\alpha=1.9\pm0.1$, which is in good agreement with the magnitude
$\alpha=1.94$, obtained using the ratio $W_3^{\rm Al}/W_3^{\rm
Be}=2.8$ from Table~\ref{tab:table6}.

Per-nucleon cross section ratio $R=(\sigma^{A_1}/A_1)/(\sigma^{A_2}/A_2)$
(Eq.~(5)) is often referred to as the transparency ratio. It is widely
believed that the target mass dependence of $R$ is determined by the
attenuation of the flux of produced particle in the nuclei which, in
turn, is governed by its in-medium width $\Gamma$ (see review in
\cite{19}).

\begin{figure}
\includegraphics[width=8.0cm]{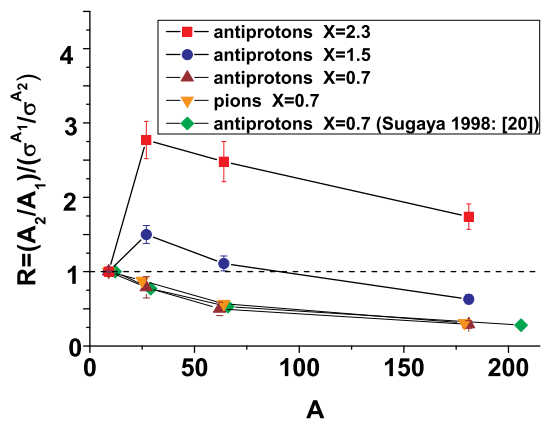}
\caption{\label{fig:fig3} (Color online) Atomic mass dependence of
the transparency ratio $R$ for antiprotons and pions in the
cumulative ($X > 1$) and noncumulative ($X \le 1$) regions.}
\end{figure}

Fig.~\ref{fig:fig3} demonstrates that the $A$-dependencies of the ratio
$R$, given in Table~\ref{tab:table5}, in the noncumulative ($X \le 1$)
and in cumulative ($X > 1$) regions are substantially different. In the
noncumulative region, the ratio $R$, normalized to the cross section on
light nucleus, is always less than 1 and it decreases as $A$ increases
for all species of secondary particles. The respective ratios of the
antiproton and $\pi^-$-meson production cross sections measured in the
present experiment in the same region, as well as the corresponding
ratios of the cross sections for antiproton creation on nuclei C, Cu and
Pb by 12~GeV protons from \cite{20}, follow the same dependence. The
measured transparency ratios for $\phi$ mesons in photon-induced
\cite{21} and proton-induced reactions \cite{22} as well as for
photoproduced $\omega$ mesons \cite{23} behave in the same way. In the
cumulative region, $R$ behaves in an essentially different way. The
values of $R(A)$ significantly exceed 1 and go up to $\sim$2.5. It must
be emphasized that the transparency ratio $R$ is determined not only by
the absorption of the produced hadrons but also by the mechanism of their
production. In the kinematically allowed region $0 \le X \le 1$, the
ratios $W_1^{\rm A}/W_1^{\rm Be}$ depend weakly on $A$ and are close to 1
(column with $j=1$ in Table ~\ref{tab:table6}). In this case the ratio
$R$ is mainly governed by the ratio of absorption factors. In the
cumulative region $X > 1$, the ratios $W_j^{\rm A}/W_j^{\rm Be}$ (columns
with $j=2$ and $j=3$ in Table~\ref{tab:table6}) significantly exceed 1
and therefore the values of $R$ depend on these two factors.

Direct comparison of the cross section ratios measured in the
$A(p,{\bar p})$ and $A(e,e')$ reactions is hindered by the use of
different targets. The ratios measured in our experiment are
$R_2(^{64}{\rm Cu}/^{9}{\rm Be})=1.11\pm0.09$ and $R_3(^{64}{\rm
Cu}/^{9}{\rm Be})=2.48\pm0.22$, whereas in the work \cite{10}
$R_2(^{56}{\rm Fe}/^{12}{\rm C})=1.17\pm0.12$ and $R_3(^{56}{\rm
Fe}/^{12}{\rm C})=1.44\pm0.18 $\footnote{Statistical and systematic
errors quoted in \cite{10} have been added in quadrature.}. However,
reasonable comparison can be performed provided that the difference
in the ratio of probabilities $W_j^{A_1}/W_j^{A_2}$ is taken into
account. According to \cite{17}, the values of $W_j^{\rm Cu}$ and
$W_j^{\rm Fe}$ (for $j=2,3$) are almost the same, while the ratio
$W_2^{\rm C}/W_2^{\rm Be} =1.5$ and $W_3^{\rm C}/W_3^{\rm Be}=2.4$
with the theoretical uncertainty of 10\%. Since $W_j^{\rm
Cu}/W_j^{\rm Be}=(W_j^{\rm Fe}/W_j^{\rm C})(W_j^{\rm C}/W_j^{\rm
Be})$, the values expected in the inclusive electron scattering are:
$R_2(^{64}{\rm Cu}/^{9}{\rm Be})=1.76\pm0.25$ and $R_3(^{64}{\rm
Cu}/^{9}{\rm Be})=3.46\pm0.55$. These values are in good agreement
with the values 1.8 and 3.4 from Table~\ref{tab:table6}. The value
$R_2(^{64}{\rm Cu}/^{9}{\rm Be})=1.32\pm0.15 $\footnote{Statistical
and systematic errors quoted in \cite{11} have been added in
quadrature. Unfortunately, the HMS data in the range of $j=3$ suffer
from poor statistics which makes the numerical comparison with our
experiment difficult.}, measured by the HMS collaboration at
$Q^2=3.73$~GeV$^2$ \cite{11}, falls between $R_2(^{64}{\rm
Cu}/^{9}{\rm Be})=1.11\pm0.09$ (obtained in our experiment) and
$1.76\pm0.25$ (the expected magnitude cited above). Therefore, the
values $R_j(A/{\rm Be})$ from the present experiment are comparable
but still less than those extracted from the data on inclusive
electron scattering off nuclei \cite{10}, \cite{11}. This difference
should be attributed to the antiproton absorption in intermediate
and heavier nuclei. If antiprotons are not absorbed in nuclei, the
ratio $f_{\rm FSI}^A/f_{\rm FSI}^{\rm Be} =1$. In this case the
ratios ${\tilde R}_j(A/{\rm Be})=R_j(A/{\rm Be})(A/9)$, shown in
Fig.~\ref{fig:fig2} by the crossed open symbols, are defined by the
probability ratios from Table~\ref{tab:table6}. The difference
between the dark and open symbols characterizes the amount of the
effect of antiproton absorption in different nuclei.

Our estimate of the antiproton absorption in nuclei is inspired by
the Quark Gluon String Model (QGSM) based on $1/N$ expansion in QCD
\cite{3}, as well as by the three-stage model of hadron formation in
nuclear medium \cite{24, 25, 26}. The model \cite{3} presumes the
existence of CBCs as an inherent property of the nuclear structure
and considers the process of cumulative hadron production $pA \to
hX$ in the reference frame where the nucleus moves with a large
momentum. This consideration is equivalent to the analysis in the
rest frame of the target nucleus. It is assumed that this nucleus is
composed of usual nucleons with a probability $W_1^A$ and of
$3j$-quark colorless CBCs with a probability $W_j^A$ where $j$
changes from 2 to $A$. According to the model, each CBC initially
contains a quark configuration of the $X$- and $P_t$-distribution.
The value $X$ determines a fraction of the quark momentum relative
to the CBC momentum in the rest frame of the moving nucleus.
Interaction of the moving nucleus with the target results in the
production of hadrons in the fragmentation range of nucleus $A$.
Cumulative hadrons with $X > 1$ originate from the quark (antiquark)
fragmentation of a massive $3j$ ($j
> 1$) CBCs into hadrons (antiprotons in our case). This model allows
quantitative description of the cross sections and some specific features
of the cumulative hadron production \cite{3}. Processes in the
fragmentation regions of the nucleus and the target become independent at
high collision energies. Indeed, experimentally observed shapes of the
cumulative hadron distributions are insensitive to the types of beam
particles. This fact, together with the observed comparability of the
cross section ratios measured in the $A(p,{\bar p})$- and $A(e,e')$-
reactions, supports the idea that the properties of the target (proton or
electron) are not essential for the analysis of the cross section ratios.
Therefore, Eqs.~(\ref{eq:three}), (\ref{eq:five}) do not include the
initial-state interaction (ISI) of the incident proton with nuclear
matter and the ratio $R$ depends only on the final-state interaction
(FSI) of the produced antiproton. The models \cite{24,25,26} consider the
hadron formation in semi-inclusive deep inelastic scattering on nuclei as
a three-step process. During the first (production) stage, the quark
(antiquark) propagates quasi-freely undergoing multiple collisions with
nucleons. During the second (formation) stage, the color neutralization
takes place and small size colorless prehadron is created which has a
reduced absorption cross section, subsequently the hadron wave function
is formed. During the third (propagation) stage, the surviving prehadron
transforms into a final state (detected) hadron. Absorption of the hadron
on its way out of the nucleus is governed by the 'normal' hadron-nucleon
cross section.

To estimate the antiproton formation length in the nuclear matter, we use
a simplified two-stage model. Our analysis does not make distinction
between the production length and formation length, dealing with the
total effect referred to as the `formation length'. During the first
(formation) stage, a quark (antiquark) ejected from CBC at some point
$O_1$ inside the nucleus propagates without interaction with the nuclear
environment until the hadronic antiproton emerges at point $O_2$. The
antiproton formation length $L$ in the nucleus rest frame is defined as
the distance traveled by the quark between points $O_1$ and $O_2$. During
the second (propagation) stage, the antiproton travels further in the
same direction before escaping the nucleus. Interactions of the
antiproton with nucleons on its way out of the nucleus are determined by
the momentum dependent total cross section $\sigma_{{\bar p}N}^{\rm
tot}(P_{\bar p})$ in the free space \cite{27}.

We evaluated the length $L$ using the following procedure. The l.h.s. of
Eq.~(\ref{eq:five}) was determined as the ratio $R$ of the measured cross
sections for each specific antiproton momentum and production angle. The
ratios $W_j^{A_1}/W_j^{A_2}$, entering the r.h.s. of Eq.~(\ref{eq:five}),
were taken from Table~\ref{tab:table6}. The ratio $f_{\rm FSI}^{A}/f_{\rm
FSI}^{\rm Be}$ depends on $L$ and characterizes the propagation stage of
a formed antiproton in the nucleus. This ratio was calculated within the
Glauber model accounting for the actual nucleon densities and arbitrary
antiproton production angles. Subsequently we treated $L$ as a free
parameter and determined the value of $L$ as satisfying
Eq.~(\ref{eq:five}). We estimated $L$ using the cross section ratios
measured in the range of dominance of CBCs with $j=3$. Calculation of the
antiproton production length using the ratio $R_3({\rm Al}/{\rm Be})$
gave $L \ge (2 \div 3)r_{\rm Al}$, where $r_{\rm Al}$ is the radius of
the Al nucleus. As already noted above, this indicates that the
absorption of antiprotons in light nuclei is weak. The values of $L$,
obtained from the experimental data on the cross section ratios $R_3({\rm
Cu}/{\rm Be})$ and $R_3({\rm Ta}/{\rm Be})$, are given in
Table~\ref{tab:table7}.

\begin{table*}
\caption{\label{tab:table7}The antiproton formation length $L$ in the
region where CBCs with $j=3$ dominate.}
\begin{ruledtabular}
\begin{tabular}[b]{|c|c|c|c|c|c|c|c|}\hline
 & & & & & & \\
$P_{\bar p}$ (GeV/$c$) & 1.72 & 1.53 & 0.92 & 0.74 & 0.66 & 0.60 & 0.58 \\
\hline
$\theta_{\bar p}$ (degree) & 59 & 59 & 97 & 97 & 119 & 97 & 119 \\
\hline $L$(Cu/Be) (fm) & $4.4_{-1.5}^{+1.6}$ & $4.4_{-1.7}^{+1.8}$&
$4.7_{- 1.8}^{+1.7}$ & $4.9_{-1.7}^{+1.8}$&
$4.9_{-1.4}^{+1.6}$ & $4.9_{-1.7}^{+1.8}$ & $5.0_{-1.8}^{+1.6}$ \\
\hline $L$(Ta/Be) (fm) & & $4.0_{-2.5}^{+1.6}$ & $4.2_{-2.4}^{+1.5}$
& $4.3_{- 2.5}^{+1.6}$ & $4.2_{-2.4}^{+1.5}$&
$4.2_{-2.4}^{+1.5}$ & $4.4_{-2.4}^{+1.5}$ \\
\hline
\end{tabular}
\end{ruledtabular}
\end{table*}

The data presented in Table~\ref{tab:table7} show that the formation
length $L$ does not depend (within the uncertanties) either on the
momentum or on the antiproton production angle and is governed only by
the properties of CBCs with $j=3$. The weighted average value of $L$,
calculated using the data from Table~\ref{tab:table7}, equals to $4.5_{-
0.7}^{+0.5}$~fm. This number is comparable to the radii of an
intermediate nuclei with $A \approx 60$ ($R_{1/2}^{\rm Cu} = 4.2$~fm).
Similar calculations, employing the ratios $R_2(A/{\rm Be})$ ($A=$ Cu,
Ta) measured in the region where CBCs with $j=2$ dominate, lead to
$L=2.8_{-0.7}^{+0.6}$~fm. In the region where CBCs with $j=1$ dominate,
$L = 2.0_{-0.8}^{+0.6}$~fm, this value being comparable with the average
separation of nucleons inside the nucleus of $1.8 \div 2$~fm.

The observed rise of $L$ with $j$ is in agreement with predictions
of the three-step model \cite{24,25,26}. According to the model, the
hadron formation length $L$ in the nuclear rest frame is
proportional to the energy of initial quark. Indeed, the energy of
initial quarks ejected from $j=1$, $j=2$, and $j=3$ CBCs increases
with $j$ since in the plateau regions the initial quarks carry the
increasing momentum fractions $0 \le X \le 1$, $1 \le X \le 2$, $2
\le X \le 3$, correspondingly. Furthermore, since the initial quark
shares its energy with produced hadrons (antiproton and nucleon in
our case to satisfy the baryon number conservation), the value of
$X$ calculated from Eq.~\ref{eq:two} using the kinematical
parameters of detected antiproton turns out to be less than the
Bjorken scaling variable $x_B$ in the inclusive electron scattering
off nuclei \cite{10,11} when the struck quark absorbs all the energy
of the virtual photon. For this reason, the plateaus observed in
cumulative antiproton production are shifted towards lower $X$
compared to those seen in \cite{10,11}.

$\pi^-$-meson production cross sections, measured in the present
experiment at the lab angles of 10$^{\circ}$ and 59$^{\circ}$
simultaneously with antiprotons, were mostly obtained in the
kinematically allowed region $X \le 1$. Study of the cumulative pion
production by 10~GeV protons on the same set of nuclear targets in
the range $1 \le X \le 3.4$ was carried out in our previous work
\cite{16}. Similar to the antiproton case, the ratios $R_j(A/{\rm
Be})$ increase with the rise of $X$, but clear plateaus in $R(X)$
were not observed. Note that the model \cite{25} implies a
flavor-dependent formation length. Contrary to a pion, an antiproton
cannot be formed by a valence quark picking up an antiquark from the
string break-up. An antiproton can only be created from new $q{\bar
q}$ pairs formed inside color strings with reduced energy resulting
from the break-up of the initial string. The string length is
getting shorter after each break, thus delaying the next pair
production \cite{28}. Formation of a $q{\bar q}$ state, i.e. pion,
takes less time than formation of a $3{\bar q}$ state, i.e.
antiproton.  Therefore, one can expect a shorter formation length of
a pion compared to that of an antiproton under similar initial
conditions. Within our simplified model, this results in significant
elongation of the second (propagation) stage for cumulative pions
and, hence, in the increase of the role of the pion final-state
interactions. These interactions may distort the observed pion
spectra and mask the plateaus.

A widely-used phenomenological description of the antiproton absorption
in nuclei boils down to determination of the effective antiproton-nucleon
cross section in nuclear matter. Our calculations within the model
accounting for both one- and two-step elementary antiproton production
processes show that this cross section lies within the range of 40--20~mb
at the antiproton momenta of 0.6-2.5~GeV/$c$, which is essentially less
than the free inelastic antiproton-nucleon cross section of 100--50~mb at
the same momenta \cite{27}. This result is consistent with the finding of
the experiment \cite{29} where strong suppression of the annihilation of
antiprotons within the nuclei was observed in proton-nucleus collisions
at the beam energy of 14--17~GeV.

\section{\label{sec4}Summary}
The inclusive antiproton and $\pi^-$-meson production cross sections were
measured at the lab angles of 10.5$^{\circ}$, 59$^{\circ}$ and in the
momentum range from 0.6 to 2.5~GeV/$c$ during interactions of 10~GeV
protons with Be, Al, Cu and Ta nuclei. Analysis of the antiproton
creation cross section ratios in the region of cumulative variable $X$
ranging from 0.5 to 2.8 was performed. It accounted both for the data
obtained in the present experiment and those previously obtained by us on
the same set of nuclear targets under lab angles of 96$^{\circ}$ and
119$^{\circ}$ and with the same initial proton energy.

It was shown that the $A$-dependence of the cumulative antiproton
production cross sections is enhanced with the increase of $X$ and is
mainly determined by the ratio of the probabilities of the existence of
CBCs with $j=1, 2, 3$ in nuclei with different mass number $A$. In the
regions where the given CBC dominates, the cross section ratios
demonstrate the presence of three plateaus. In the region of each plateau
these ratios exhibit the scaling behavior, i.e. they do not depend
neither on $X$ nor on $P_t^2$. Observation of the plateau in $A(p,{\bar
p})$- and in $A(e,e')$-reactions \cite{10,11} in the kinematically
forbidden region is indicative for the presence of compact baryon
configurations in nuclei---objects of nuclear structure existing at
small-distance scales. It was found that the magnitudes of the ratios
$R_j(A_1/A_2)$ measured in the $A(p,{\bar p})$-reaction are comparable
with those observed in the $A(e,e')$-reaction, which evidences for a weak
absorption of cumulative antiprotons in nuclear medium. In the regions
where contribution to the antiproton production cross section from CBCs
with $j=1, 2, 3$ dominates, the antiproton formation length $L$ increases
with the rise of $j$ and reaches the value of 4.5~fm for $j=3$, which is
commensurable with the radius of the Cu nucleus.

Study of cumulative hadron production in proton-induced nuclear reactions
provides new information on the features of the hadronization process in
the cold nuclear medium and may also aid interpretation of such an
observable as jet quenching in the hot medium, which was found at RHIC in
high energy heavy-ion collisions \cite{30} and will be studied at much
higher energies at the LHC.

\begin{acknowledgments}
The authors gratefully acknowledge stimulating discussions with A.B.
Kaidalov and O.V. Kancheli.
\end{acknowledgments}


\begin{thebibliography}{0}%
\makeatletter
\providecommand \@ifxundefined [1]{%
 \@ifx{#1\undefined}
}%
\providecommand \@ifnum [1]{%
 \ifnum #1\expandafter \@firstoftwo
 \else \expandafter \@secondoftwo
 \fi
}%
\providecommand \@ifx [1]{%
 \ifx #1\expandafter \@firstoftwo
 \else \expandafter \@secondoftwo
 \fi
}%
\providecommand \natexlab [1]{#1}%
\providecommand \enquote  [1]{``#1''}%
\providecommand \bibnamefont  [1]{#1}%
\providecommand \bibfnamefont [1]{#1}%
\providecommand \citenamefont [1]{#1}%
\providecommand \href@noop [0]{\@secondoftwo}%
\providecommand \href [0]{\begingroup \@sanitize@url \@href}%
\providecommand \@href[1]{\@@startlink{#1}\@@href}%
\providecommand \@@href[1]{\endgroup#1\@@endlink}%
\providecommand \@sanitize@url [0]{\catcode `\\12\catcode `\$12\catcode
  `\&12\catcode `\#12\catcode `\^12\catcode `\_12\catcode `\%12\relax}%
\providecommand \@@startlink[1]{}%
\providecommand \@@endlink[0]{}%
\providecommand \url  [0]{\begingroup\@sanitize@url \@url }%
\providecommand \@url [1]{\endgroup\@href {#1}{\urlprefix }}%
\providecommand \urlprefix  [0]{URL }%
\providecommand \Eprint [0]{\href }%
\providecommand \doibase [0]{http://dx.doi.org/}%
\providecommand \selectlanguage [0]{\@gobble}%
\providecommand \bibinfo  [0]{\@secondoftwo}%
\providecommand \bibfield  [0]{\@secondoftwo}%
\providecommand \translation [1]{[#1]}%
\providecommand \BibitemOpen [0]{}%
\providecommand \bibitemStop [0]{}%
\providecommand \bibitemNoStop [0]{.\EOS\space}%
\providecommand \EOS [0]{\spacefactor3000\relax}%
\providecommand \BibitemShut  [1]{\csname bibitem#1\endcsname}%
\let\auto@bib@innerbib\@empty
\end{thebibliography}%


\begin{thebibliography}{99}
\bibitem{1} M. M. Sargsian {\it et al.}, J. Phys. G: Nucl. Part. Phys.
{\bf 29}, R1 (2003).
\bibitem{2} S. V. Boyarinov {\it et al.}, Yad. Fiz. {\bf 50}, 1605
(1989).
\bibitem{3} A. V. Efremov {\it et al.}, Phys. Atom. Nucl. {\bf 57}, 874
(1994).
\bibitem{4} L. L. Frankfurt and M. I. Strikman, Phys. Rep. {\bf 76}, 215
(1991);\\
 L. L. Frankfurt and M. I. Strikman, Phys. Rep. {\bf 160}, 235 (1998).
\bibitem{5} C. Ciofi degli Atti and S. Simula, Phys. Rev. C {\bf 53},
1689 (1996).
\bibitem{6} G. Berlard, A. Dar and G. Eilam, Phys. Rev. D {\bf 22}, 1547
(1980).
\bibitem{7} M. A. Braun and V. V. Vechernin, Nucl. Phys. B {\bf 427}, 614
(1994).
\bibitem{8} H. J. Pirner and J. P. Vary, Nucl. Phys. A {\bf 358}, 413c
(1981).
\bibitem{9} D. B. Day {\it et al.}, Phys. Rev. Lett. {\bf 59}, 427
(1987).
\bibitem{10} K. Egiyan {\it et al.}, Phys. Rev. C {\bf 68}, 014313 (2003)
;\\
 K. Egiyan {\it et al.}, Phys. Rev. Lett. {\bf 96}, 082501 (2006).
\bibitem{11} N. Fomin, arXiv: nucl-ex/0812.2144.
\bibitem{12} J. Aubert {\it et al.}, Phys. Lett. B {\bf 123}, 275 (1983).
\bibitem{13} J. Gomez {\it et al.}, Phys. Rev. D {\bf 49}, 4348 (1994).
\bibitem{14} J. Seely {\it et al.}, Phys. Rev. Lett. {\bf 103}, 202301
(2009) .
\bibitem{15} L. B. Weinstein {\it et al.}, Phys. Rev. Lett. {\bf 106},
052301 (2011).
\bibitem{16} S. V. Boyarinov {\it et al.}, Phys. Atom. Nucl. {\bf 57},
1379 (1994);\\
 ibid {\bf 54}, 71 (1991); {\bf 56}, 72 (1993).
\bibitem{17} M. Sato {\it et al.}, Phys. Rev. C {\bf 33}, 1062 (1986).
\bibitem{18} J. W. Cronin {\it et al.}, Phys. Rev. D {\bf 11}, 3105
(1975);\\
 C. Bromberg {\it et al.}, Phys. Rev. Lett. {\bf 42}, 1202 (1979).
\bibitem{19} S. Leupold, V. Metag, U. Mosel, Int. J. Mod. Phys. E {\bf
19}, 147 (2010).
\bibitem{20} Y. Sugaya {\it et al.}, Nucl. Phys. A {\bf 634}, 115 (1998).
\bibitem{21} T. Ishikawa {\it et al.}, Phys. Lett. B {\bf 608}, 215
(2005).
\bibitem{22} A. Polyanskiy {\it et al.}, Phys. Lett. B {\bf 695}, 74
(2011).
\bibitem{23} M. Kotulla {\it et al.}, Phys. Rev. Lett. {\bf 100}, 192302
(2009).
\bibitem{24} B. Z. Kopeliovich {\it et al.}, Nucl. Phys. A {\bf 782},
224c (2007).
\bibitem{25} S. Domdey {\it et al.}, Nucl. Phys. A {\bf 825}, 200 (2009).
\bibitem{26} W. K. Brooks {\it et al.}, J. Phys. Conf. Ser. 299: 012011,
2011.
\bibitem{27} Review of Particle Properties, Phys. Rev. D {\bf 50}, 1336
(1994).
\bibitem{28} B. Kopeliovich and A. H. Rezaeian, Int. J. Mod. Phys. E {\bf
18}, 1629 (2009).
\bibitem{29} I. Chemakin {\it et al.}, Phys. Rev. C {\bf 64}, 064908
(2001).
\bibitem{30} J. Adams {\it et al.}, Phys. Rev. Lett. {\bf 91}, 172302
(2003);\\
 S. S. Adler {\it et al.}, Phys. Rev.C {\bf 69}, 034910 (2004).
\end{thebibliography}

\end{document}